\documentclass[final,onecolumn,prfluids,aps,amsfonts,amssymb,showpacs,floatfix,nofootinbib,superscriptaddress]{revtex4-2}
\usepackage{amsmath}
\usepackage{graphicx}
\usepackage[dvipsnames]{color}
\usepackage{hyperref}
\usepackage{psfrag}
\usepackage{stmaryrd}
\usepackage{amssymb}
\usepackage{float}
\usepackage{xcolor}
\usepackage{comment}
\usepackage[T1]{fontenc}
\usepackage[french,english]{babel}
\usepackage[utf8]{inputenc}
\usepackage{lmodern}
\usepackage{multirow,makecell}
\usepackage{subcaption} 

\usepackage{geometry}
\newcommand{\diff}{\mathrm{d}}

\begin{document}

\title{Laboratory observation of internal gravity wave turbulence in a three-dimensional, large-scale facility}

\author{Nicolas Lanchon}
\affiliation{Universit\'e Paris-Saclay, CNRS, FAST, 91405 Orsay, France}
\affiliation{Universit\'e Paris-Saclay, CEA, CNRS, SPEC, 91191 Gif-sur-Yvette, France}
\author{Samuel Boury}
\affiliation{Université Paris Cité, CNRS, MSC, 75013 Paris, France}
\affiliation{Universit\'e Paris-Saclay, CNRS, FAST, 91405 Orsay, France}
\author{Pierre-Philippe~Cortet}
\email[]{pierre-philippe.cortet@universite-paris-saclay.fr}
\affiliation{Universit\'e Paris-Saclay, CNRS, FAST, 91405 Orsay, France}

\date{\today}

\begin{abstract}
    The search for solutions to the theory of weakly non-linear internal gravity wave turbulence is an active research topic. It is notably stimulated by the fact that this regime could drive fine-scale ocean dynamics for which the identification of a physical model could yield improved parametrizations in global oceanic models. In this context, analytical works lead to diverse predictions and the experimental observation of a regime of developed weakly-non-linear internal wave turbulence constitutes a major, still unachieved, objective of experimentalists in the field. In this study, building on recent experimental developments, we present laboratory observations of internal gravity wave turbulence in a linearly stratified fluid, performed in a large-scale, three-dimensional facility allowing the forcing of long-wavelength internal waves. Our setup allows to access large Reynolds numbers favoring the development of turbulent power-law spectra while keeping the Froude number relatively low in order to remain weakly non-linear. As the forcing amplitude increases, the flow seems to approach a wave turbulence regime: we indeed observe the progressive construction of a continuous distribution of energy both in the frequency and wave number spaces, whereas the spatio-temporal spectra indicate that the energy remains almost exclusively carried by internal gravity waves verifying the dispersion relation. We finally show that, as the transition to turbulence proceeds, the bicoherence spectrum of the velocity field becomes smooth over the internal wave frequency domain, taking values of the order of the Froude number. While these observations are in line with the phenomenology of weakly non-linear wave turbulence, the power laws in $k^{-3}$ we report over about a decade for the horizontal and vertical spatial energy spectra agree with the prediction that can be made from raw dimensional arguments for a strongly non-linear ``saturated wave'' turbulence. Whether these power laws could alternatively be compatible with a weakly non-linear wave turbulence regime remains to be explored theoretically.
\end{abstract}

\maketitle

	\section{Introduction}
	\label{sec:sec1}

Density stratified fluids are the support of a specific class of waves, dispersive and anisotropic, called internal gravity waves. Considering an inviscid and linearly stratified fluid, their dispersion relation writes
	\begin{equation}\label{eq:disp}
		\omega = N \frac{k_\perp}{\sqrt{k_\perp^2 + k_z^2}},
	\end{equation}
where $\omega$ is the angular frequency, and $k_\perp$ and $k_z$ are the norm of the horizontal and vertical components of the wave vector $\mathbf{k}$, respectively. The strength of the stratification is quantified in Eq.~(\ref{eq:disp}) by the buoyancy frequency $N=\sqrt{-g/\rho_0 \, \diff\bar{\rho}/\diff z}$, where $g$ is the gravitational acceleration, $\rho_0$ the mean density of the fluid, and $\bar{\rho}(z)$
the vertical profile of the density of the fluid at rest ($z$ is oriented opposite to gravity). It is worth noting that Eq.~(\ref{eq:disp}) is obtained from the Euler equation under the Boussinesq approximation, which consists in considering weak density variations with respect to the reference density $\rho_0$~\cite{Staquet2002,Sutherland2010,Dauxois2018}.
	
Ubiquitous in geophysical flows, internal gravity waves contribute to the redistribution of energy over frequencies and spatial scales through a variety of non-linear processes~\cite{McComas1977} and, as such, they play a key role in the ``small-scale'' turbulent dynamics of the atmosphere and the oceans~\cite{Pedlosky1987, Wunsch2004, Vallis2006, Riley2008, MacKinnon2017}. In the atmosphere, a strongly non-linear regime of internal wave turbulence is observed~\cite{Nastrom1985,Dewan1986,Cot2001} for which the ``strongly stratified turbulence'' phenomenology has been proposed~\cite{Brethouwer2007}. On the contrary, at small scales in the oceans ---typically from a few hundred meters to a few meters in the vertical direction~\cite{Polzin2011}--- a weakly non-linear regime called internal wave turbulence is expected~\cite{Dematteis2024}. The small scales in question are not resolved in global oceanic models~\cite{MacKinnon2017, Stensrud2007, Polzin2014, Gregg2018} and are accounted for by empirical parametrizations of the power drained by the small-scale turbulence. In this context, the development of parametrizations based on a physical model could provide a major contribution to improving oceanic dynamics predictions~\cite{Dematteis2024}.
	
From a fundamental point of view (and considering Péclet numbers much larger than $1$ for which the effects of the diffusion of mass are negligible~\cite{Caulfield2021}), the different regimes of turbulence in a stratified fluid~\cite{Davidson2013,Riley2012,Cortet2024} can be classified by introducing three non-dimensional numbers: the Reynolds number $Re = \tau_\nu/\tau_{\rm nl}$, the Froude number $Fr = 1/\tau_{\rm nl}N$, and the non-dimensional frequency $\omega^*= \omega/N$, where $\tau_{\rm nl}$ and $\tau_\nu$ are the characteristic non-linear and viscous timescales of the flow structures at scale $\ell$, respectively, and $1/\omega$ is their linear timescale (the wave period), which might be different from $\tau_{\rm nl}$ (with $\omega\tau_{\rm nl} \geq 1$). In practice, a cautious analysis of the anisotropic equations of the dynamics is necessary to explicit the scaling of the non-linear and viscous times as a function of the horizontal $u_\perp$ and vertical $u_z$ velocity components and of the horizontal $\ell_\perp$ and vertical $\ell_z$ length scales. In this context, in order to simplify the problem, theoreticians most often consider the strongly anisotropic limit $\ell_z \ll \ell_\perp$ in which the non-linear and viscous timescales follow $\tau_{\rm nl} \sim \ell_\perp/u_\perp$ and $\tau_{\nu} \sim \ell_z^2/\nu$, respectively~\cite{Brethouwer2007,Lanchon2023b}. On the contrary, most experimental works involve waves with $\omega^*=\omega/N$ neither too close to $0$ nor too close to $1$, for which $\ell \sim \ell_z \sim \ell_\perp$ and simple scaling can be written for the Reynolds, $Re=u\ell/\nu$, and Froude, $Fr=u/N\ell$, numbers.

Considering high Reynolds numbers, two families of turbulence can be identified. A first one, the strongly non-linear regime also called the ``critical balance'' or ``saturated wave'' regime, is expected when $Fr \simeq \omega ^*$, reflecting the fact that the linear and non-linear timescales are of the same order of magnitude at a given spatial scale. Drawing on a raw dimensional analysis, one expects in this regime $1$D kinetic energy spectra scaling as ${E(k)\sim N^2k^{-3}}$~\cite{Staquet2007,Nazarenko2011b}. A more thorough study, using the additional assumption of strong anisotropy $k_\perp \ll k_z$, leads to the ``strongly stratified turbulence'' phenomenology~\cite{Brethouwer2007} describing a regime which is also referred to as ``Layered Anisotropic Stratified Turbulence'' (LAST) in the literature~\cite{Caulfield2021}. In this regime, that has been proposed as the relevant scenario for the intermediate and small-scale atmospheric turbulence~\cite{Riley2008}, distinct predictions for the horizontal and vertical $1$D spatial energy spectra have been made, ${E(k_\perp)\sim \varepsilon^{2/3} k_\perp^{-5/3}}$ and ${E(k_z)\sim N^2 k_z^{-3}}$, respectively (where $\varepsilon$ is the mean rate of energy dissipation per mass unit).

The other type of stratified turbulence, called ``wave turbulence''~\cite{Nazarenko2011, Galtier2022, Cortet2024}, is expected when the Froude number $Fr$ is much smaller than the non-dimensional frequency $\omega^*$. Here, the kinetic and potential energies are carried by quasi linear internal gravity waves and exchanged among spatial scales over timescales much larger than the wave period. Non-linearities remain weak compared to the effects of the buoyancy, hence the other name of ``weak turbulence'' used to refer to this regime. As already said, weak stratified turbulence has often been suggested as a potential explanation for the oceanic dynamics at small scales \cite{McComas1977,McComas1981,Polzin2011,Dematteis2024} without, however, a definitive confirmation so far. The general weak/wave turbulence formalism~\cite{Nazarenko2011, Galtier2022} has been applied to the case of stratified fluids by several theoretical teams in the past decades leading to various (anisotropic) predictions for the energy spectra~\cite{Pelinovsky1977,Caillol2000,Lvov2001,Lvov2004,Lvov2010,Lanchon2023b,Dematteis2024,Labarre2024,Shavit2025}. The obtained solution depends in particular on the assumptions made during the derivation (two-dimensionality, low frequency waves, selection of a certain type of interactions, exclusion of the steady mode), the general case seeming out of reach analytically. In this context, numerical simulations of the so-called kinetic equation ---the central equation in wave turbulence theory--- appears as a natural strategy to identify the solutions of the weak internal gravity wave turbulence problem. Although technical obstacles to evaluate the collision integral (driving the dynamics in the kinetic equation) have long been overcome~\cite{McComas1981}, this approach, very costly in computational resources, has only recently started to yield promising results~\cite{Scott2024,Labarre2025}.

An alternative strategy is attempting to reach the regime of weak turbulence of internal waves in laboratory experiments or numerical experiments (Direct Numerical Simulations). This is, however, also very challenging in both cases. On the one hand, it requires large facilities in the experimental case to fulfill simultaneously the conditions, $Re=u\ell/\nu \gg 1$ and $Fr=u/N\ell \ll \omega^* \leq 1$, under which weak turbulence is expected. On the other hand, it begs for important computational resources to numerically simulate the Navier-Stokes equations in this regime, especially because of the linear/non-linear timescale separation, characteristic of wave turbulence, which imposes an integration of the dynamics over particularly long durations. Satisfying the second condition (of weak non-linearity) $Fr \ll \omega^* \leq 1$ is indeed equivalent to having separated linear and non-linear timescales. In practice, this condition involves to force internal waves, by imposing independently their frequency and their velocity. This strategy has been explored in several experimental works~\cite{Benielli1996, Benielli1998, Brouzet2016, Brouzet2017, Davis2020, Savaro2020, Rodda2022, Rodda2023, Lanchon2023} as well as in numerical simulations of the Navier-Stokes equations under the Boussinesq approximation~\cite{LeReun2018}.

A detailed review of these works is presented in Ref.~\cite{Cortet2024}. We will therefore limit ourselves here to briefly describing some of their important common features. In all these studies, internal gravity waves are forced at a specific frequency $\omega_0^*$ either directly by oscillating an object in the fluid domain~\cite{Brouzet2016, Brouzet2017, Davis2020, Savaro2020, Rodda2022, Rodda2023, Lanchon2023} or indirectly, through a parametric instability, by oscillating the fluid container at $2\omega_0^*$~\cite{Benielli1996,Benielli1998,LeReun2018}. Most of them report, when increasing the forcing amplitude on the road to turbulence, the emergence of a first non-linear state of the flow characterized by the presence of discrete peaks in the temporal energy spectrum, most often associated to internal wave eigenmodes (or quasi-eigenmodes) of the experimental domain~\cite{Benielli1996,Benielli1998,Brouzet2016,Brouzet2017,Davis2020,Savaro2020,Lanchon2023}. The prevalence of these peaks clearly suggests that the attraction of energy by eigenfrequencies of the experimental aquarium is central to such studies attempting to reach internal wave turbulence. This notably implies a discretization of the energy in the frequency and wave vector spaces which strongly deviates these (almost turbulent) flows from the regimes described by theories as well as from the small-scale oceanic turbulence, where finite size effects are either absent or weak, and a turbulent cascade forms an energy continuum. In several studies discussed in this paragraph, further increasing the forcing amplitude leads the temporal energy spectrum to progressively become continuous over the internal wave frequency domain~\cite{Brouzet2016,Brouzet2017,Savaro2020,Davis2020,Lanchon2023}. This transition from discrete to continuous nevertheless seems to drive the flow toward a strongly non-linear regime. Indeed, a common feature of most of the works discussed in this paragraph is the emergence at the largest amplitudes of forcing of $1$D spatial energy spectra following scalings in $k_\perp^{-3}$ and $k_z^{-3}$ (or $k^{-3}$ for the spectrum averaged over the orientation)~\cite{LeReun2018,Davis2020,Lanchon2023} or of a $\omega^{-3}$ scaling for the temporal spectrum interpreted as the transposition of a $k_z^{-3}$ spatial spectrum by an intense turbulent sweeping~\cite{Benielli1996}. These scalings suggest that the flow approaches a regime of strongly non-linear ``saturated wave'' turbulence, which is also supported in some of these studies by the observation of intense wave breaking events~\cite{Rodda2023} and of a significant mixing of the background density stratification~\cite{Brouzet2016,Brouzet2017}. We should nonetheless notice that these $k^{-3}$ power-law behaviors are not observed over more than half-a-decade of wave numbers, revealing that the turbulent cascade is still not well developed at the Reynolds numbers reached in these experiments.

Two main features should thus be improved in laboratory experiments aiming at reaching a fully developed, weakly non-linear, internal wave turbulence regime. First, the concentration of energy in the wave eigenmodes of the fluid domain should be prevented. In this regard, the work of Lanchon~\textit{et al.}~\cite{Lanchon2023} proposed an efficient practical solution consisting in introducing slightly inclined panels at the top and/or at the bottom of the experimental domain: by slightly modifying the wavelength of the reflected waves, the panels prevent the appearance of standing modes in the flow. The second change that should be implemented is a significant increase in the wavelength $\ell$ at which the energy is injected and, incidentally, an increase in the size of the water tank. The goal is to access large Reynolds numbers $Re=u\ell/\nu$, to observe turbulent spectra with developed power laws, while remaining in the weakly non-linear regime, which implies reducing the Froude number of the flow $Fr=u/N\ell$. The study presented in this article specifically aims at meeting these needs. In the following, we report results obtained with an experimental device allowing the forcing of long-wavelength internal gravity waves in a $2$~m-high $2$~m-large three-dimensional water tank, filled with $8\,000$ liters of a stratified column of salt water.

	\section{Experimental setup}
	\label{sec:setup}

\begin{figure}[!htbp]
		\centerline{\includegraphics[width = 10cm]{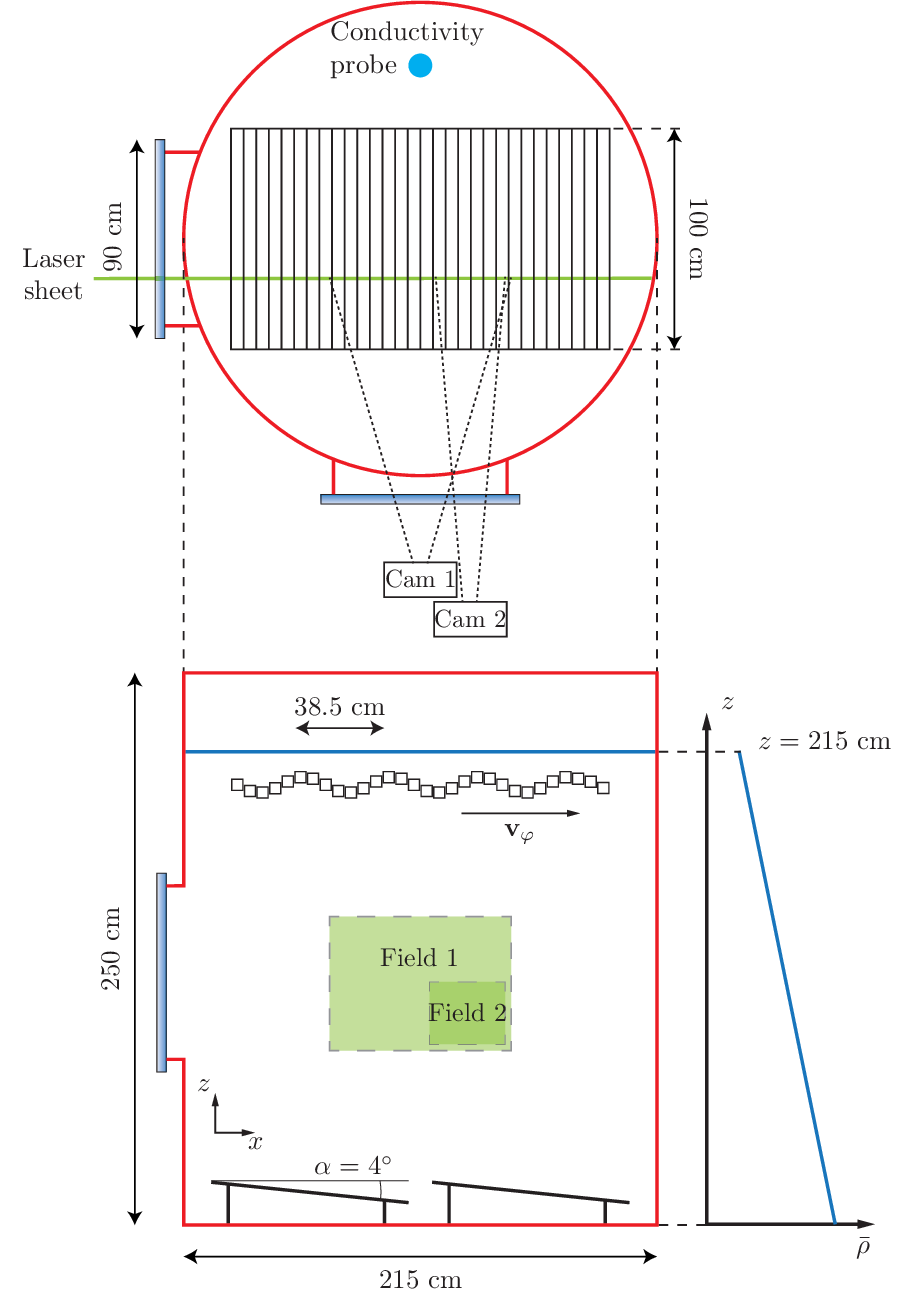}}
		\caption{Schematic of the top and side views of the experimental apparatus. A horizontal wave maker, composed of $30$ vertically oscillating bars, produces a sinusoidal horizontally propagating transverse wave motion of phase velocity $\mathbf{v}_\varphi$. The cylindrical tank is filled with a linearly stratified water solution of salt up to a height of $215$~cm.} \label{fig:fig1}
	\end{figure}

\begin{figure}[!htbp]
		\centering
		\includegraphics[width=0.9\textwidth]{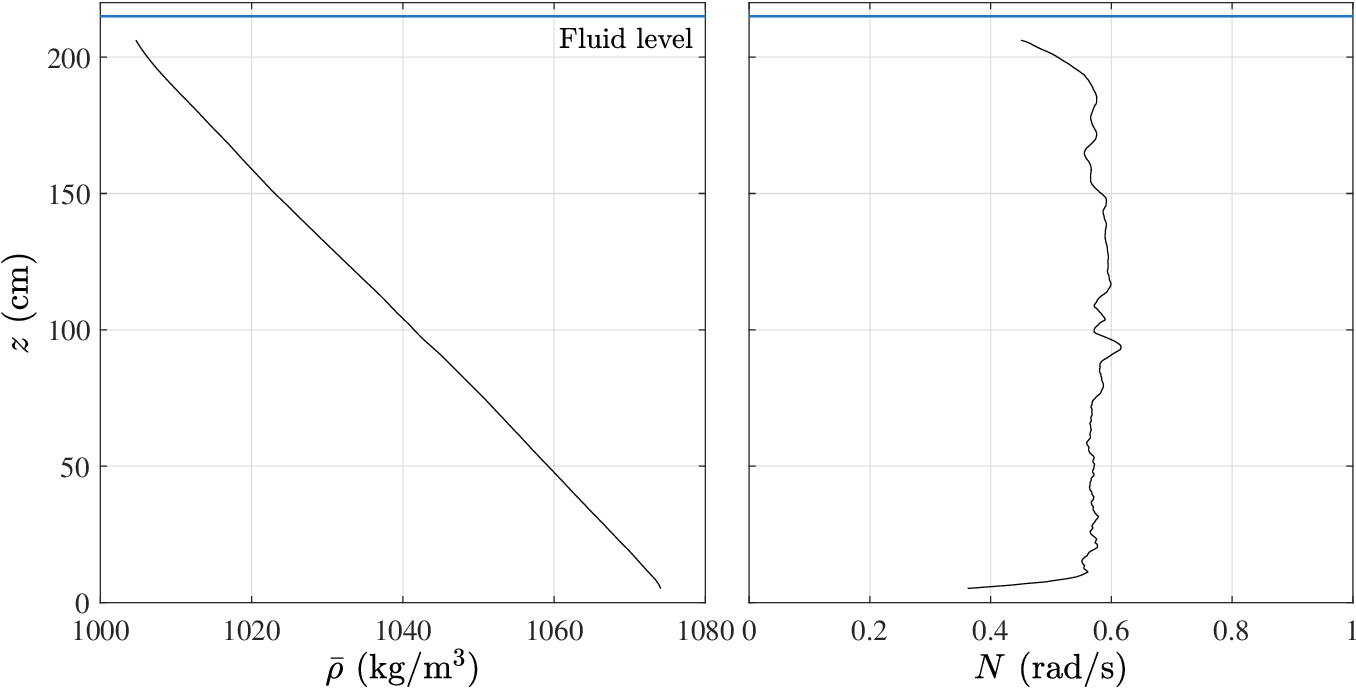}
		\caption{Left: Vertical density profile $\bar{\rho}(z)$ measured after the tank filling procedure. The reference $z=0$ corresponds to the bottom of the experimental tank. Right: Corresponding profile of buoyancy frequency $N(z)= \sqrt{-g/\rho_0 \, d\bar{\rho}/dz}$. The buoyancy frequency measured over the water column (outside of the top and bottom boundary layers) evolves in the range $0.58\pm 0.02$~rad/s. Its average value over the region where the PIV measurements are realized, at mid-depth of the fluid, is $N=0.59$~rad/s.}
		\label{fig:fig2}
	\end{figure}

	Experiments are performed in a cylindrical tank of diameter $2.15$~m and height $2.5$~m equipped with two square plexiglas vertical openings of $90\, \times\, 90$~cm$^2$ for visualization purposes, as illustrated in Fig.~\ref{fig:fig1}. The tank is filled up to a height of $2.15$~m with a linearly stably stratified water solution of salt obtained thanks to the double-bucket method~\cite{Fortuin1960, Oster1963, Hill2002}. Two geared pumps are used to connect the two $4\,500$~L filling tanks to the $9\,000$~L experimental tank. The stratification is made over $21$~h using $500$~kg of salt, and we obtain a typical buoyancy frequency $N$ of about $0.6$~rad/s. The density profile $\bar{\rho}(z)$ (of the fluid at rest) is measured thanks to a conductivity and temperature probe (Mettler Toledo InLab 731-ISM-10m) mounted on a motorized vertical translation axis and calibrated using a fluid densimeter (Anton Paar DMA35). A typical stratification is presented in Fig.~\ref{fig:fig2} with (left panel) the measured density profile $\bar{\rho}(z)$ and (right panel) the corresponding buoyancy frequency profile $N(z)$. While processing the data, we retain the value $N=0.59$~rad/s consistently measured, over the different experiments, in the bulk of the fluid (in the region where the velocity field is measured, see Fig.~\ref{fig:fig1}).

	The flow is forced at the top of the fluid domain by means of a horizontal wave generator, adapted from \cite{Brunet2019}, consisting of a series of $30$ horizontal bars of $5\,\times\, 5$~cm$^2$ square section, $1$~m length, spaced by $0.5$~cm gaps, and aligned along a diameter of the tank (see Fig.~\ref{fig:fig1}). Each bar is connected to a linear motor able to drive it in a vertical motion. The wave generator motion approximates a truncated sinus wave profile of spatial period $\lambda_0=7\times 5.5$~cm~$=38.5$~cm ($5.5$~cm is the width of an oscillating bar plus the gap separating two bars) and spatial extension about $4.3\,\lambda_0$, explicitly
	\begin{equation}
		Z(x,t) = H + A \left[ \cos (\omega_0 t - k_0 x) - 1 \right],
		\label{eq:forcage}
	\end{equation}
	with $k_0 = 2 \pi / \lambda_0$, and $\omega_0=0.81\,N$ the forcing angular frequency. The amplitude $A$ of the bars vertical motion is varied between $2$ and $18$~mm. In the linear regime, this wave generator is expected to produce an internal wave beam at frequency $\omega_0$, which is therefore propagating in a direction making an angle $\theta=\sin^{-1}(\omega_0/N) \simeq 54^\circ$ with the horizontal. The forced wavelength is expected to be $\lambda_f=\lambda_0 \sin\theta \simeq 31$~cm. In addition, two $90\,\times\, 90$~cm$^2$ square panels inclined at an angle $\alpha\simeq 4^\circ$ with respect to the horizontal are placed at the bottom of the tank (see Fig.~\ref{fig:fig1}) in order to inhibit the formation of standing wave modes in the fluid domain, following a method described in Ref.~\citep{Lanchon2023}.

	The two components ($u_x,u_z$) of the velocity field are measured in a vertical area in the central region of the tank using a two-dimension two-component Particle Image Velocimetry system (PIV). We nevertheless use two cameras aiming at two (overlapping) fields of view of different sizes (see Fig.~\ref{fig:fig1}) in order to access to an extended range of spatial scales during the data analysis. The fluid is seeded with $10~\mu$m glass tracer particles, injected at different heights of the stratification thanks to a homemade immersed manifold. The PIV vertical laser plane, shifted by $30\mathrm{~cm}$ from the central vertical axis of the tank, is created using a $140$~mJ Nd:YAG pulsed laser. Successive images covering two areas of $\Delta x \times \Delta z = 817 \times 615$~mm$^2$ (field $1$ in Fig.~\ref{fig:fig1}) and $\Delta x \times \Delta z = 310\, \times\, 261$~mm$^2$ (field $2$) are recorded by two cameras, of $2360\, \times\, 1776$~pixels and $2432\, \times\, 2048$~pixels, respectively. PIV cross-correlation between successive images are computed using $24\, \times\, 24$~pixels interrogation windows with a $50\%$ overlap, providing velocity fields with a spatial resolution of $4.2$~mm/pixel and $1.5$~mm/pixel, respectively. The image acquisition rate is adjusted depending on the flow typical velocity, ranging from $2$ to $5$~Hz. Image acquisition starts $30$ forcing periods $T= 2\pi/\omega_0$ before the forcing device is started, and lasts between $530\,T$ and $1030\,T$ in total, depending on the experimental run.
	
	\begin{table}[!htbp]
		\begin{center}
		\begin{tabular}{c c c c}
		\hline
		~~$A$ (mm)~~ & ~~$u_{\mathrm{rms}}$ (mm/s)~~ & ~~$Re_{\mathrm{rms}}$~~ & ~~$Fr_{\mathrm{rms}}$~~ \\
		\hline \hline
		2 & 1.1 & 340 & 0.006 \\
		3 & 1.4 & 430 & 0.008 
		\\
		6 & 2.4 & 740 & 0.013 \\
		8 & 3.0 & 930 & 0.016 \\
		10 & 3.7 & 1100 & 0.020 \\
		14 & 4.6 & 1400 & 0.025 \\
		18 & 5.4 & 1700 & 0.030 \\
		\hline
		\end{tabular}
		\caption{Parameters of the experiments presented in the article. $A$ is the amplitude of the vertical motion of the bars of the wave generator (see Eq.~\eqref{eq:forcage}). The reported Reynolds $Re_{\mathrm{rms}} = u_{\mathrm{rms}} \lambda_f/\nu$ and Froude $Fr_{\mathrm{rms}} = u_{\mathrm{rms}}/\lambda_f N$ numbers are estimated using measurements of the flow rms velocity $u_{\mathrm{rms}}$, the forced wavelength $\lambda_f = \lambda_0 \omega_0/N \simeq 31$~cm and $\nu = 10^{-6}$~m$^{2}$/s for the fluid kinematic viscosity.}
		\label{tab:experiments}
		\end{center}
	\end{table}

	The parameters of the experiments discussed in the following are presented in Tab.~\ref{tab:experiments}. The root mean square (rms) velocity $u_{\mathrm{rms}}$ used to estimate the Reynolds $Re_{\mathrm{rms}}$ and Froude $Fr_{\mathrm{rms}}$ numbers is computed as $u_{\mathrm{rms}} = \left\langle \sqrt{\left\langle u_x^2 + u_z^2 \right\rangle_t}\right\rangle_{\mathbf{x}}$, where $\left\langle~\right\rangle_t$ stands for the temporal average (in the statistically steady state of the flow) and $\left\langle~\right\rangle_{\mathbf{x}}$ for the spatial average over the PIV region (large view, field $1$).

	\section{Results}
	\label{sec:sec3}
	
\begin{figure}[!htbp]
\centering
\includegraphics[width=0.9\textwidth]{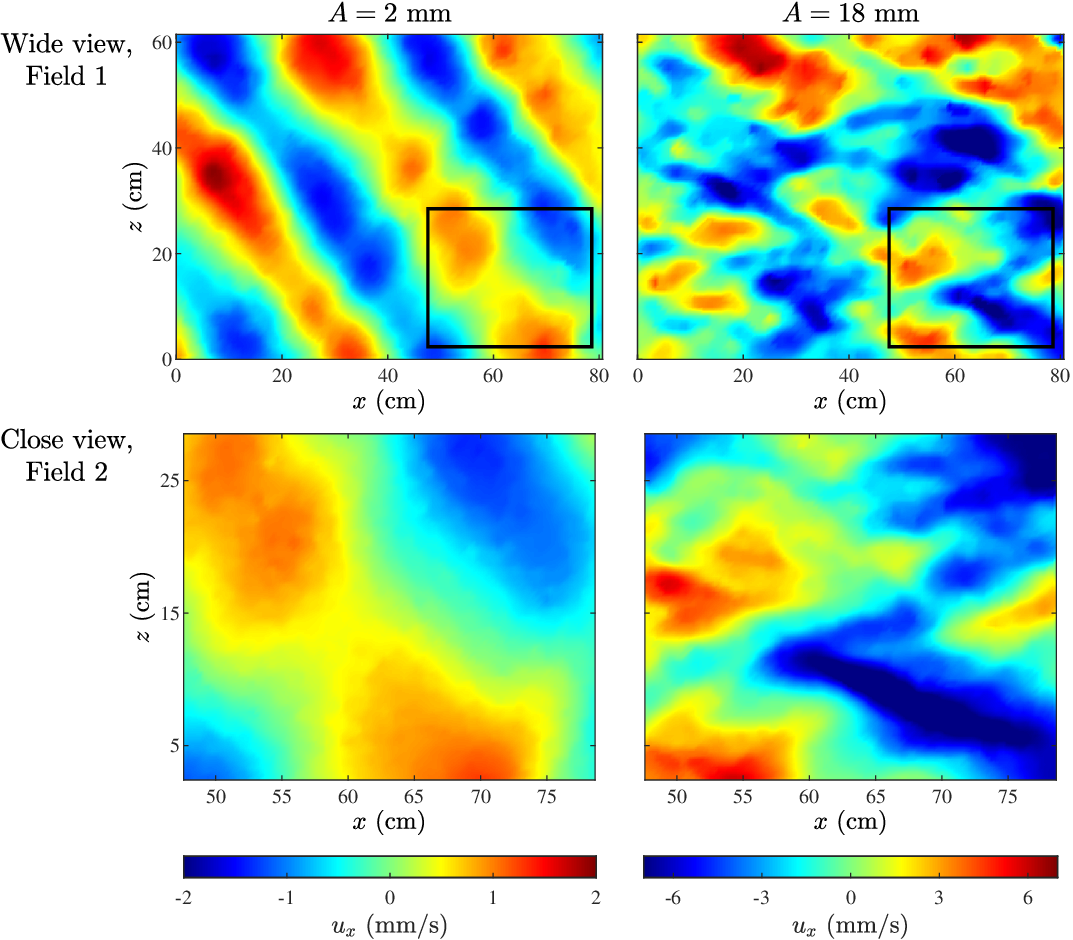}
\caption{Snapshots of the horizontal component of the velocity field, at $t = 250\,T$ after the start of the wave generator, computed from images of the wide (top panels) and narrow (bottom panels) view cameras. The fields in close view correspond to the rectangular region indicated at the bottom right corner of the large view fields. The left panels correspond to the experiment at $A=2$~mm forcing amplitude (linear regime) and the right panels to the experiment at $A=18$~mm (turbulent regime).} \label{fig:velocityFields}
\end{figure}
	
	First, we present in Fig.~\ref{fig:velocityFields} snapshots of the horizontal component of the velocity field computed from images of the narrow and wide view cameras. The velocity fields are reported for the experiments at $2$~mm and $18$~mm forcing amplitude which correspond to a flow in a linear and a turbulent regime, respectively. In the linear regime at low forcing amplitude (Fig.~\ref{fig:velocityFields}, left), the plane wave structure of the forced mode can be clearly identified with a typical horizontal wavelength perfectly consistent with the generator wavelength ${\lambda_0=38.5}$~cm. The inclination of the planes of constant phase is also consistent with the tilt of $\theta=\sin^{-1}(\omega_0/N) \simeq 54^\circ$ with respect to the horizontal expected from the dispersion relation. The field observed in Fig.~\ref{fig:velocityFields}(left) nevertheless shows some deviations from a pure plane wave structure which results from the reflections of the forced wave on the tank walls interfering with the primary beam. At large forcing amplitude $A=18$~mm, the velocity field snapshots (Fig.~\ref{fig:velocityFields}, right) reveal a flow composed of a wide spectrum of scales smaller than the forced wavelength, with the qualitative appearance of wave turbulence.
	
    \begin{figure}[!htbp]
	    \centering
	    \includegraphics[width=0.9\textwidth]{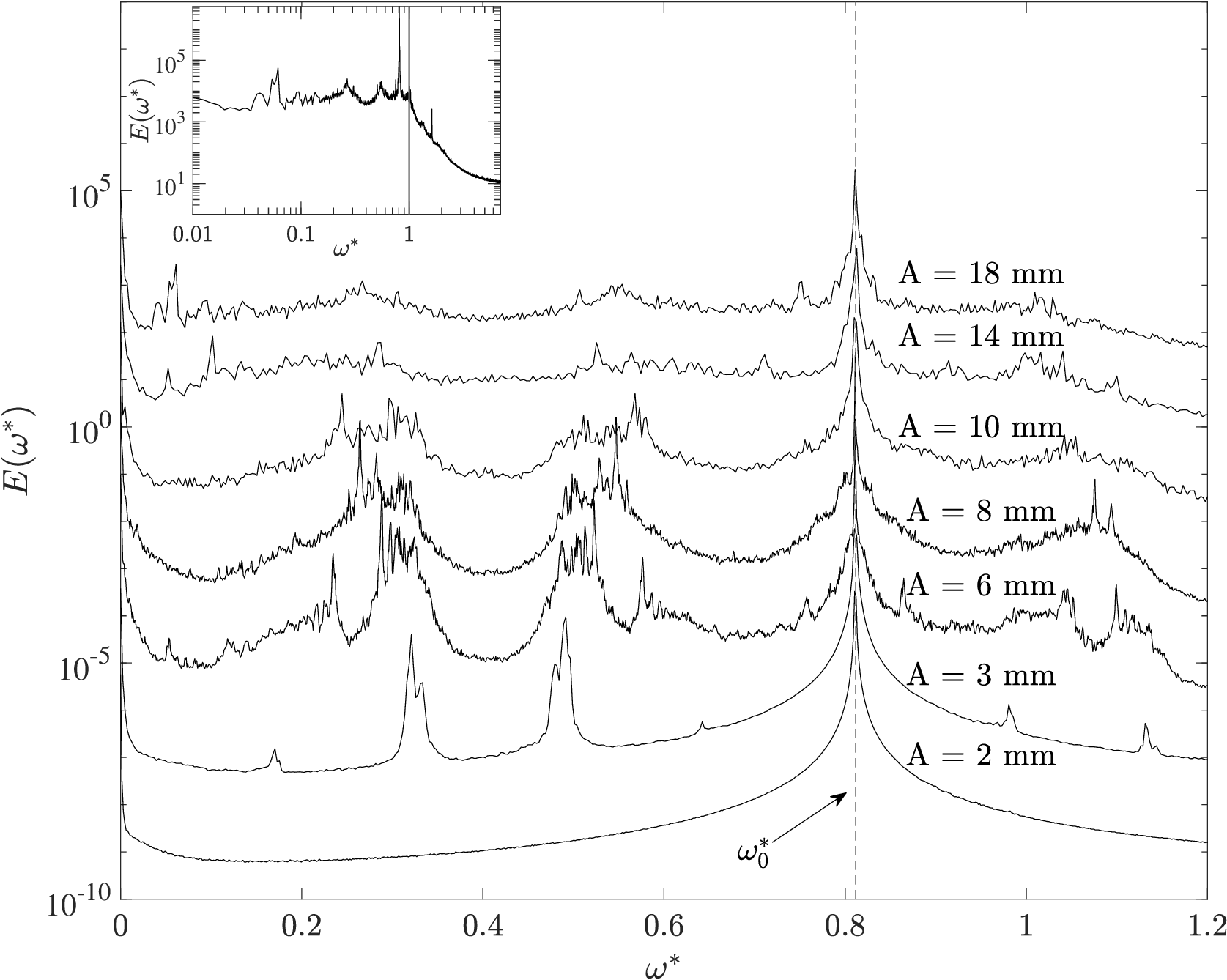}
	    \caption{Temporal kinetic energy spectrum $E(\omega^*)$ as a function of the non-dimensional frequency $\omega^*=\omega/N$ computed for all the experiments at various forcing amplitudes. The spectra are computed over $330$ to $530$ forcing periods $2\pi/\omega_0$ in the statistically steady state of the experiments. In all spectra, we observe a dominant energy peak at the forcing frequency $\omega_0^* = 0.81$ (vertical dashed line). A vertical shift by a factor $30$ has been introduced between successive spectra for better visualization. (Insert) Same temporal kinetic energy spectrum for $A = 18$~mm in log-log scale. The vertical line indicates $\omega^* = 1$.}
	    \label{fig:specTemp}
	\end{figure}

In Fig.~\ref{fig:specTemp}, we explore in a more quantitative way the generated flows by reporting the temporal kinetic energy spectra computed from the PIV measurements (large view, field~$1$) for the set of experiments listed in Tab.~\ref{tab:experiments}.
This series of curves illustrates the transition of the flow from a linear to a turbulent regime as the forcing amplitude goes from $A=2$~mm to $A=18$~mm. The spectrum of the experiment at $A=2$~mm reveals a single energy peak at the forcing frequency $\omega_0^*=0.81$, confirming the linear nature of the flow (if we set aside a very weakly energetic peak at zero frequency). Triadic Resonance Instability (TRI) arises in the experiment at $A=3$~mm, with two clear subharmonic peaks at $\omega^*_1 \simeq 0.32$ and $\omega^*_2\simeq 0.49$ in triadic resonance with the primary wave frequency~\cite{Dauxois2018}. These energy peaks are accompanied by much weaker peaks (two orders of magnitude lower than the peaks at $\omega^*_1$ and $\omega^*_2$) at frequencies in non-linear coupling with the three leading modes (e.g., $\omega^*_2-\omega^*_1$, $2(\omega^*_2-\omega^*_1)$). These modes might be the trace of the production of bound waves as noticed in Refs.~\cite{Rodda2022,Lanchon2023}. As the forcing amplitude increases, a progressive filling of the temporal energy spectrum is observed, mainly at subharmonic frequencies, i.e. $\omega \leq \omega_0$. At $A=6$~mm, we indeed observe that the two subharmonic peaks of the experiment at $A=3$~mm give way to two couples of wide subharmonic bumps in triadic temporal resonance with the forced mode (the bumps are symmetric with respect to half the forcing frequency $\omega_0/2$).
Then, as the forcing amplitude is increased up to $A=18$~mm, the subharmonic energy bumps progressively transform into a nearly flat continuum of energy over the whole internal wave frequency domain $\omega^*\leq 1$. For the experiment at the largest forcing amplitude $A=18$~mm, as evidenced by the insert in Fig.~\ref{fig:specTemp}, the energy spectrum decays quite rapidly above the buoyancy frequency $N$ (i.e. above $\omega^*=1$), which is the cutoff frequency for internal waves, suggesting that the flow might have reached a kind of wave turbulence regime.

In several previous experimental studies of internal gravity waves, the transition of the flow towards a turbulent regime as the forcing amplitude is increased was shown to go through a regime where the temporal spectrum is dominated by an ensemble of sharp peaks due to the concentration of energy in resonance frequencies associated to wave eigenmodes of the fluid domain~\cite{Savaro2020,Davis2020,Lanchon2023}. Lanchon~\textit{et al.}~\cite{Lanchon2023} demonstrated that introducing slightly tilted panels at the top and the bottom of the fluid domain allows to inhibit this concentration process. In the present experiments where we also implemented such tilted panels, we do not observe the emergence of a family of sharp energetic peaks in the series of temporal energy spectra reported in Fig.~\ref{fig:specTemp} (even if a few small peaks are observed here and there), suggesting that, at all studied forcing amplitudes, the flow is mainly composed of propagating internal waves.

Finally, it is important to note the presence in the energy spectra of Fig.~\ref{fig:specTemp} of a peak of energy at zero frequency, which corresponds to a nearly steady mode of the flow. This mode has already been reported in previous studies of internal wave turbulence~\cite{Rodda2022,Lanchon2023}, where it was shown, in a given horizontal slice, to be close to a large vortex of vertical axis, nearly centered in the water tank and of the scale of the water tank. The rate of this fluid rotation was shown to slowly evolve with the vertical coordinate, even involving changes in the direction of rotation depending on the height. We recover the same structure in the present experiments as can be seen in the velocity field reported in Appendix~\ref{app:slowmode} for the experiment at $A=8$~mm. It is interesting to note that the energy in the peak at zero frequency in the spectra of Fig.~\ref{fig:specTemp} increases from weak values (less than $1\%$) to about $15\%$ of the total energy of the flow as the forcing amplitude goes from $A=2$~mm to $18$~mm.

A dedicated study would be necessary to properly identify the non-linear mechanism at the origin of this mean flow. Although it is beyond the scope of this article, we can nevertheless suggest that it could result from a non-linear process directly affecting the primary wave mode at the forcing frequency. Such processes have indeed often been reported in stratified fluid experiments and in numerical simulations where internal waves are directly driven by a harmonic forcing. In these studies, the non-linear process may occur either in the bulk of the forced wave mode~\cite{Sutherland2006,Bordes2012b,Semin2016,Jamin2021} or in the vicinity of the wave makers~\cite{Ermanyuk2008,King2009,Fan2020}. In some of these works, the non-linear mechanism has been identified to be of streaming type~\cite{Bordes2012b,Semin2016,Fan2020,Jamin2021} or of Stokes drift type~\cite{Sutherland2006}. These two non-linear processes take place in oscillating flows with a spatially inhomogeneous amplitude and result in the production of a steady flow proportional to the square of the periodic base flow velocity~\cite{Sutherland2006,Bordes2012b,Semin2016,Fan2018}. Finally, it is worth mentioning that there might be some connection between the emergence of a slow horizontal shear flow observed in our experiments and the one reported in experiments aiming at mimicking the quasi-biennial oscillation of the Earth equatorial stratosphere. In these experiments~\cite{Plumb1978,Semin2024}, a slow horizontal flow, non-linearly driven by internal gravity waves, undergoes periodic reversals at a frequency far lower than that of the forced internal waves.

	\begin{figure}[!htbp]
	    \centering
	    \includegraphics[width=\textwidth]{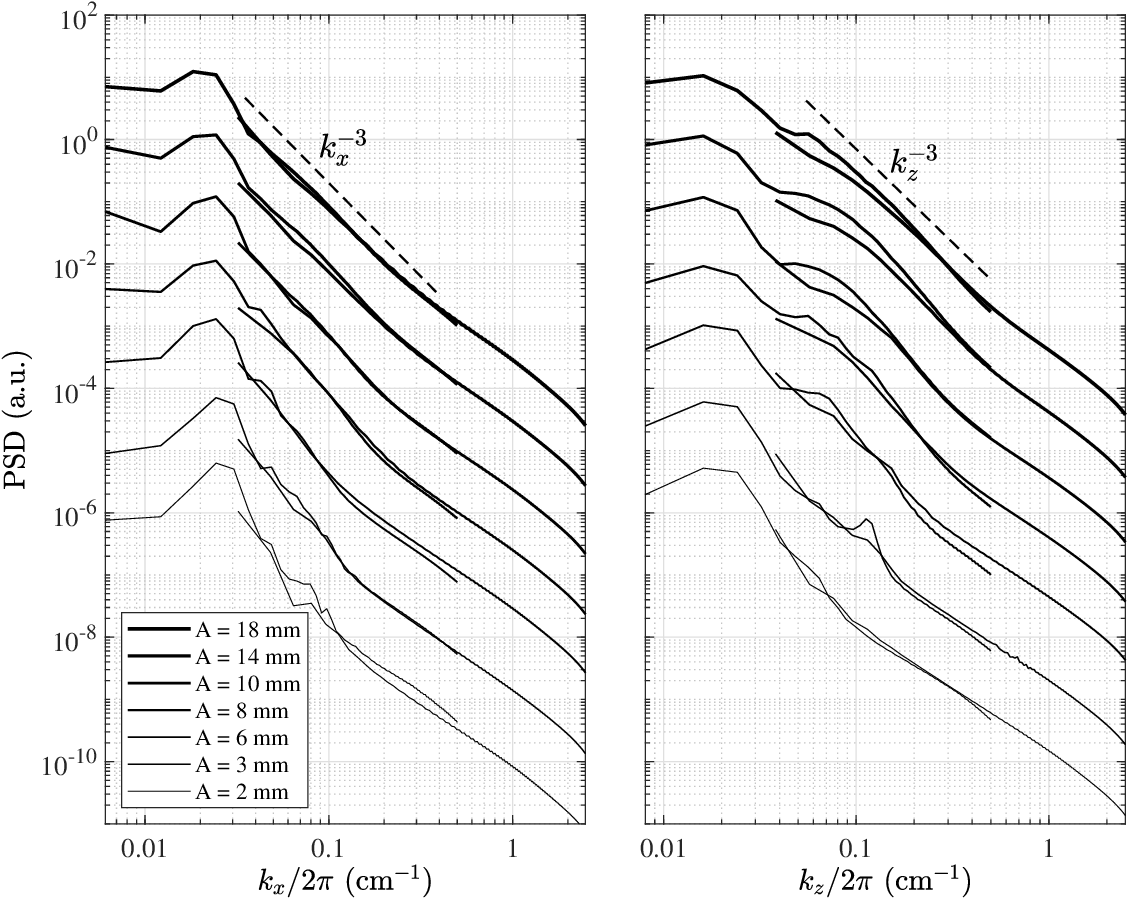}
	    \caption{One dimensional spatial kinetic energy spectra in the horizontal (left) and vertical (right) directions computed in the stationary regime for experiments at various forcing amplitudes. For the sake of clarity, a vertical shift by a factor $5$ is introduced between successive spectra. For each spectrum, the data from both cameras, large view and close view, are reported.}
	    \label{fig:specSpat}
	\end{figure}

In Fig.~\ref{fig:specSpat}, we now explore the spatial content of the flow by reporting the $1$D kinetic energy spectra in the horizontal (left panel) and vertical (right panel) directions for the experiments listed in Tab.~\ref{tab:experiments}. These spectra are estimated by computing the $1$D spatial Fourier transform of the instantaneous two-point velocity correlation, along $x$ or $z$, using the Wiener-Khinchin theorem, before taking the temporal average and spatial average over the remaining spatial direction (for more details, see Appendix~A in Ref.~\cite{Lanchon2023}). The power spectral densities obtained from both cameras, with different sizes of field of view, are shown in order to extend the accessible range of wave numbers. For a given experiment, if we set aside the boundaries for each of the two spectra, we generally observe a good agreement over their overlapping intervals both in $k_x$ and $k_z$. Focusing first on the horizontal spectra (Fig.~\ref{fig:specSpat}, left), we observe a dominant bump of energy around $k_x/2\pi = 0.025$~cm$^{-1}$, a wave number consistent with the horizontal wavelength $\lambda_0=38.5$~cm prescribed by the wave generator to the forced mode. This bump of energy at large scale, initially alone for the linear experiment at small forcing amplitude $A=2$~mm (lowest curve), is progressively complemented by additional energy spots at larger wave numbers as the forcing amplitude is increased. This non-linear transition eventually leads, for the largest forcing amplitude $A=18$~mm (top curve), to a continuum of energy following a power law in $k_x^{-3}$, starting just above the forcing energy bump and spanning over nearly a decade of wave numbers, from $0.035$~cm$^{-1}$ to $0.25$~cm$^{-1}$. Above $k_x/2\pi \simeq 0.25$~cm$^{-1}$, a slower decay of the energy spectrum at $A=18$~mm is observed and corresponds a priori to wave numbers in the noise of the PIV measurements. This behavior can actually be observed in all the spectra of Fig.~\ref{fig:specSpat}(left) at large wave numbers. It is therefore worth noting that the (steep) power law in $k_x^{-3}$ observed in the spatial energy spectrum at $A=18$~mm possibly extends to wave numbers larger than $0.25$~cm$^{-1}$, at which scales our current PIV measurement noise would nevertheless prevent from seeing it.

Looking now at the vertical spectra of Fig.~\ref{fig:specSpat}(right), we observe a dominant bump of energy at large scale around $k_z/2\pi =0.015$-$0.020$~cm$^{-1}$, which is again compatible with the vertical wavelength, $\lambda_0 \tan(\theta)\simeq 53$~cm, expected for the forced mode. As in the horizontal spectra, a continuum of energy progressively emerges at wave numbers larger than the forced wave number as the forcing amplitude increases. Note that this continuum, however, starts at a wave number significantly larger than the forcing one (typically three times). For the experiment at the largest forcing amplitude $A=18$~mm, this continuum in the spatial spectrum seems compatible with a power law in $k_z^{-3}$ extending from $k_z/2\pi \simeq 0.06$~cm$^{-1}$ to $0.4$~cm$^{-1}$. We should highlight that this last remark is mainly based on the spectrum computed from the large view camera. The spectrum computed from the close-view camera indeed deviates here significantly from the other one at its smallest wave numbers (in the range $k_z/2\pi \simeq 0.05$~cm$^{-1}$ to $0.15$~cm$^{-1}$). 

As noticed in the introduction, $1$D spatial kinetic energy spectra compatible with a power law in $k_x^{-3}$, $k_z^{-3}$ and/or $k^{-3}$ (for the angular averaged spectrum) have already been reported in several experimental and numerical studies~\cite{Benielli1996,LeReun2018,Davis2020,Lanchon2023} in which energy was injected into internal gravity waves in order to reach a turbulent regime. These $k^{-3}$ power-laws were, however, not observed over more than half-a-decade of wave numbers. Here, we clearly observe, for our experiment at the largest forcing amplitude $A=18$~mm, a $k_x^{-3}$ scaling for the $1$D horizontal energy spectrum over nearly a decade. The same result is also observed for the vertical energy spectrum with a $k_z^{-3}$ behavior over nearly a decade. This last result (for the vertical spectrum) should nevertheless be taken with some caution since there are some discrepancies between the vertical spectra measured with the data from our large-view and close-view cameras. As also discussed in the introduction, such scaling in $k^{-3}$ are compatible with regimes of ``saturated wave/critical balance'' turbulence, in which the non-linear time saturates on the value of the wave period and $1$D kinetic energy spectra in $N^2k^{-3}$ are expected~\cite{Staquet2007,Nazarenko2011b,Cortet2024}.

\begin{figure}[!htbp]
    \centering
    \includegraphics[width=12 cm]{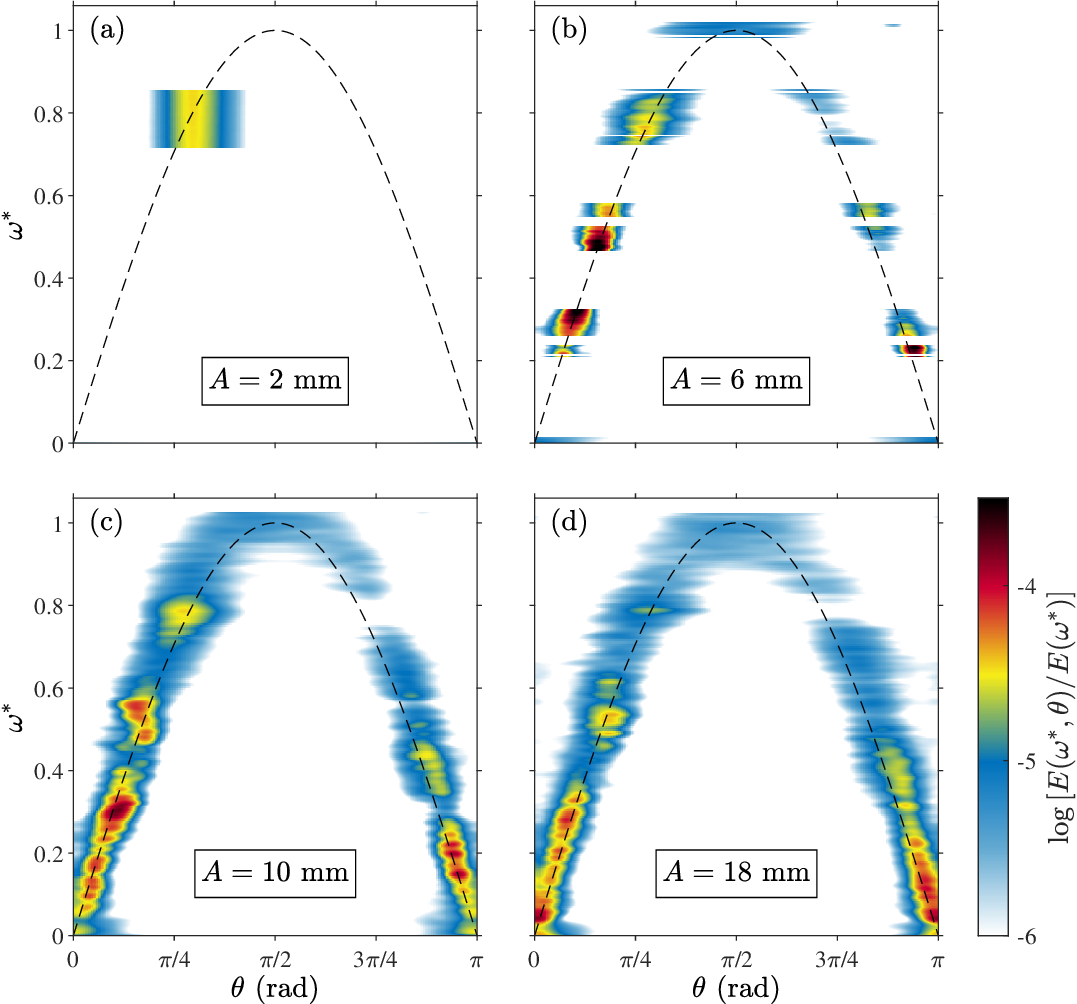}
    \caption{Normalized energy content $\log\left[E(\omega^*,\theta)/E(\omega^*)\right]$ of the measured velocity field in the $(\omega^*,\theta=\tan^{-1}(k_x/k_z))$ plane for the experiments at forcing amplitude $A=2$, $6$, $10$ and $18$~mm. A threshold is applied so that data are reported only when $E(\omega^*) \geq 10^{-4} \times E(\omega^*_0)$. The dashed curve shows the dispersion relation $\omega^* = \sin(\theta)=k_x/\sqrt{k_x^2+k_z^2}$ of internal gravity waves propagating in the measurement plane (i.e. waves with $k_y=0$).}
    \label{fig:omegaVsTheta}
\end{figure}

After studying separately the spatial and temporal energy spectra of the flow, it is important to analyze its spatio-temporal content in order to assess the prevalence of internal gravity waves. In Fig.~\ref{fig:omegaVsTheta}, we show the normalized energy content $\log\left[E(\omega^*,\theta)/E(\omega^*)\right]$ of the measured velocity field as a function of the non-dimensional angular frequency $\omega^*$
and of the angle $\theta=\tan^{-1}(k_x/k_z)$ of the wave vector with the vertical in the measurement plane. The energy spectrum $E(\omega^*,\theta)$, reported for the experiments at forcing amplitudes $A=2$, $6$, $10$ and $18$~mm, is computed from the integration of the spatio-temporal spectrum $E(\omega^*,k_x,k_z)$ (representations of which are shown in Appendix~\ref{app:spatiotemp}) along the direction defined by the angle $\theta=\tan^{-1}(k_x/k_z)$ as follows
\begin{equation}\label{eq:spectre_theta}
	    E(\omega^*,\theta) = \int_{-\infty}^\infty E(\omega^*,k \cos{\theta},k \sin{\theta})\,k dk\,.
\end{equation}
In such a representation, the energy of internal gravity waves propagating in the vertical measurement plane (i.e. waves with $k_y=0$) should gather along the internal wave dispersion relation $\omega^*=\sin(\theta)$ which is reported as a dashed line in the panels of Fig.~\ref{fig:omegaVsTheta}. It is interesting to note that, in the representation of Fig.~\ref{fig:omegaVsTheta} following from the fact we study PIV measurements in a given vertical plane, the concentration of energy along the line $\omega^*=\sin(\theta)=k_x/\sqrt{k_x^2+k_z^2}$ is also fully compatible with a flow composed of a statistically axisymmetric distribution of internal gravity waves around the vertical axis in the Fourier space (see Appendix~B of Ref.~\cite{Lanchon2023}).

For the linear experiment at forcing amplitude $A=2$~mm (Fig.~\ref{fig:omegaVsTheta}a), we observe a single spot of energy lying on the dispersion relation at the forcing frequency $\omega_0^*=0.81$. When increasing the forcing amplitude, the onset of the TRI previously observed in the temporal spectrum (Fig.~\ref{fig:specTemp}) leads to the emergence of energy spots at discrete subharmonic frequencies again along the dispersion relation of internal waves (Fig.~\ref{fig:omegaVsTheta}b). Then, as the forcing amplitude is further increased, the progressive enrichment of the temporal energy content of the flow over the wave frequency domain ($\omega^*\leq 1$) observed in Fig.~\ref{fig:specTemp} is associated in Fig.~\ref{fig:omegaVsTheta} with the emergence of a continuum of energy gathered along nearly the whole wave dispersion curve (Figs.~\ref{fig:omegaVsTheta}c and \ref{fig:omegaVsTheta}d). Overall, the data reported in Fig.~\ref{fig:omegaVsTheta} indicates that, from the linear regime at $A=2$~mm up to the turbulent regime at $A=18$~mm, the flows produced in our experiments are almost only composed of internal gravity waves verifying the wave dispersion relation. 

As previously mentioned, and as demonstrated in Appendix~B of Ref.~\cite{Lanchon2023}, the observable reported in Fig.~\ref{fig:omegaVsTheta} does not allow to discriminate between a two-dimensional ($2$D) wave turbulence invariant along the horizontal out-of-PIV-plane direction $y$ and a three-dimensional ($3$D) turbulence with a statistically axisymmetric distribution of wave vectors. We nevertheless believe that the turbulent flow produced in our experiment is far from being two-dimensional. While the geometry of the forcing device imposes that the primary wave beam is nearly $2$D (i.e. invariant in the horizontal $y$ direction), most of the modes composing the turbulent flow are more likely to be $3$D for several reasons. On the one hand, triadic resonant interactions of internal gravity waves are known to drive energy transfers in three dimensions and spontaneously produce wave modes propagating in different vertical planes~\cite{McComas1977,Ghaemsaidi2019,Mora2021,Labarre2024}. On the other hand, the multiple reflections of the waves, including the forced wave beam, on the vertical cylindrical wall of the experimental apparatus will also rapidly increase the statistical axial symmetry of the flow. As such, the wave turbulence we report may not be perfectly statistically axisymmetric, but it is certainly far from being two-dimensional due to the non-linear mechanisms at play and the geometry of the system.

To conclude our discussion of Fig.~\ref{fig:omegaVsTheta}, it is finally interesting to note that the nice agreement between the spatio-temporal energy spectra and the dispersion relation we observe demonstrates that the mean horizontal shear flow discussed earlier does not induce any significant Doppler shift in wave frequency. As observed in previous experimental works on internal gravity~\cite{Rodda2023} and inertial~\cite{Campagne2015} waves, the advection of an internal wave turbulence by a large-scale horizontal slow mode can indeed profoundly modify its frequency signature. Figure~\ref{fig:omegaVsTheta} shows that the horizontal mean flow here is, however, too weak to produce such an effect. We should also highlight that, even when the Doppler shift/sweeping effect by a horizontal mean flow is weak, some theoretical works (see, e.g., Ref.~\cite{Kafiabad2019}) predict that it can produce a scattering of internal waves leading to a diffusion of wave energy toward small spatial scales. This process, however, leads to horizontal spatial energy spectra scaling in $k_x^{-2}$ and a conservation of the wave frequency, two predictions which are not in line with our observations.

    In a weakly non-linear internal wave system, the turbulent dynamics is expected to be driven by interactions involving triads of waves verifying, in addition to the dispersion relation, spatial and temporal resonances on their wave vectors (${\bf k_1}\pm{\bf k_2}\pm{\bf k_3}={\bf 0}$) and frequencies ($\omega_1 \pm \omega_2 \pm \omega_3=0$), respectively~\cite{McComas1977,Dauxois2018}. Since the spatial resonance is intrinsic to the description of the Euler equation in the Fourier space~\cite{Sagaut2009,Nazarenko2011,Galtier2022}, it is crucial to assess the importance of the triadic temporal resonances in our flow. To this end, we compute the bicoherence spectrum of the horizontal velocity field, defined as
	\begin{equation}
		B(\omega_1,\omega_2)  = \frac{| \langle \tilde{u}_x(\mathbf{x},\omega_1)\tilde{u}_x(\mathbf{x},\omega_2)\tilde{u}_x^\star(\mathbf{x},\omega_1+\omega_2) \rangle_{\mathbf{x}}|}{\sqrt{e(\omega_1)e(\omega_2)e(\omega_1+\omega_2)}},
		\label{eq:bispectre}
	\end{equation}
	where $\tilde{u}_x(\mathbf{x},\omega)$ is the temporal Fourier transform of the horizontal velocity, the exponent $^\star$ denotes the complex conjugate, $\left\langle~\right\rangle_{\mathbf{x}}$ stands for the spatial average over the measurement region, and  $e(\omega)=\langle|\tilde{u}_x(\mathbf{x},\omega) |^2 \rangle_{\mathbf{x}}$.
	This bicoherence spectrum, already considered in a few experimental studies of internal waves~\cite{Brouzet2016,Rodda2022}, probes the phase correlation of modes at frequencies $\omega_1$, $\omega_2$, and $\omega_1+\omega_2$, and therefore the strength of their triadic coupling. For an internal wave turbulence matching the classical theoretical assumptions of large domain and weak non-linearity ($Fr\ll\omega^*$), the bicoherence spectrum $B(\omega_1,\omega_2)$ is expected to be a smooth function that takes values of the order of the non-linearity parameter $Fr/\omega^*$ over the wave frequency domain~\cite{Monsalve2020,Hasselmann1963}. It is instead expected to be composed of peaks with values of order $1$ when an ensemble of discrete modes in the frequency and wave vector space interact non-linearly (see the discussion in Ref.~\cite{Monsalve2020}).

\begin{figure}[!htbp]
    \centering
    \includegraphics[width=12 cm]{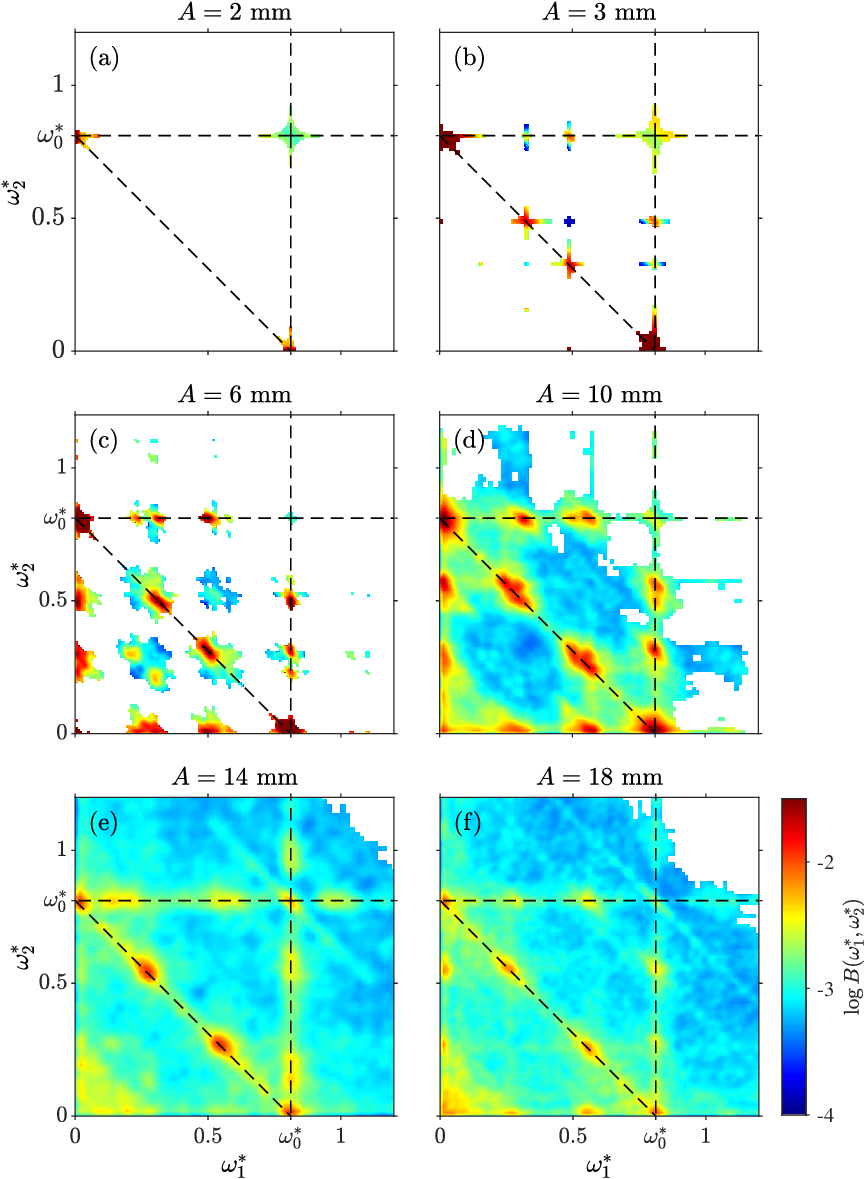}
    \caption{Logarithm of the bicoherence $B$ as a function of $(\omega_1^*,\omega_2^*)$ and for forcing amplitudes $A=2$, $3$, $6$, $10$, $14$, and $18$~mm. The value of the bicoherence is reported only when $\sqrt{e(\omega_1)e(\omega_2)e(\omega_1+\omega_2)} \geq 10^{-5} \times \max(\sqrt{e(\omega_1)e(\omega_2)e(\omega_1+\omega_2)})$ in order to focus on significantly energetic interactions. In each panel, the dashed lines indicate the forcing frequency, i.e. $\omega_1^*=\omega_0^*$ and $\omega_2^*=\omega_0^*$, and couples of frequencies in resonance with the forced mode such that $\omega_1^* + \omega_2^* = \omega_0^*$.}
    \label{fig:bicoherence}
\end{figure}

	The bicoherences $B$ associated to the experiments at forcing amplitude $A=2$, $3$, $6$, $10$, $14$, and $18$~mm are presented as a function of $(\omega_1^*,\omega_2^*)$ in Fig.~\ref{fig:bicoherence}. In the (quasi-)linear regime ($A=2$~mm, Fig.~\ref{fig:bicoherence}a), three peaks can be observed, associated to the non-linear interaction of the forced mode at $\omega_0^*$ with itself and the mean flow at zero frequency, in line with the corresponding temporal spectrum in Fig.~\ref{fig:specTemp}. Such bicoherence spectrum is consistent with a (weak) steady flow directly forced by non-linearities affecting the sinusoidal waves at the forcing frequency. Such kind of process have often been reported in harmonically forced stratified fluid experiments and numerical simulations~\cite{Sutherland2006,Ermanyuk2008,King2009,Bordes2012b,Jamin2021,Fan2020}, where in some cases the authors were even able to identify the non-linear mechanism at play to be steady streaming~\cite{Bordes2012b,Fan2020,Jamin2021} or Stokes drift~\cite{Sutherland2006}.
	
	Figure~\ref{fig:bicoherence}(b), corresponding to the experiment at forcing amplitude $A=3$~mm, shows the emergence of additional isolated spots of bicoherence mainly at 
	($\omega_1^* \simeq 0.32, \omega_2^* \simeq 0.49$) and ($\omega_1^* \simeq 0.49, \omega_2^* \simeq 0.32$) revealing a discrete triadic resonance with the forced mode such that $\omega_1^* + \omega_2^* \simeq \omega_0^*$. These two spots are in line with the two peaks of energy observed in the temporal spectrum at $A=3$~mm in Fig.~\ref{fig:specTemp} which have been attributed to the triadic resonance instability of the forced waves at frequency $\omega_0^*$. In Fig.~\ref{fig:bicoherence}(b), weaker spots of bicoherence can also be observed along the lines $\omega_1^*=\omega_0^*$ and $\omega_2^*=\omega_0^*$, revealing secondary interactions between the forced mode and the modes produced by TRI.

	Increasing the forcing amplitude to $A=6$~mm (Fig.~\ref{fig:bicoherence}c), the discrete hierarchical pattern of bicoherence spots becomes richer with the emergence of new generations of discrete triadic interactions (in a way similar to Ref.~\cite{Brouzet2016}). We also note that the spots of bicoherence start to spread in the frequency space $(\omega_1^*,\omega_2^*)$. In line with this feature, exploring larger forcing amplitudes $A=10$, $14$, and $18$~mm (Figs.~\ref{fig:bicoherence}d, e, and f, respectively) leads the bicoherence to progressively evolve from a discrete pattern of spots of large amplitude (of order $10^{-1}$) to a smooth function taking values in the range $10^{-3}$ to $10^{-2}$ in the wave frequency domain (and smaller values outside). These values of the bicoherence for the experiment at the largest forcing amplitude are roughly of the order of the Froude number $Fr$ reported in Table~\ref{tab:experiments} as expected for an internal wave turbulence satisfying the random phase approximation~\cite{Monsalve2020}. Overall, the series of bicoherence spectra shown in Fig.~\ref{fig:bicoherence} reveals a gradual transition from a regime of discrete-wave interactions to a regime of internal wave turbulence, characterized by smooth temporal, spatial, and bicoherence spectra.
	
\section{Conclusion}
\label{sec:sec4}

The present study is a substantial addition to previous works seeking to achieve a fully-developed weakly-non-linear internal gravity wave turbulence in a laboratory experiment. By upscaling the volume of the fluid domain to $8\,000$~liters while keeping it three-dimensional, and by increasing the wavelength of the forced internal wave, we manage to reach a Reynolds number $Re$ (based on the rms velocity and the actual wavelength) of about $2\,000$ while keeping the Froude number $Fr$ relatively small (to $0.03$), which is the theoretical condition for the wave dynamics to remain weakly non-linear.

By gradually increasing the forcing amplitude, we observe a transition from a quasi-linear regime to what arguably constitutes an internal wave turbulence. This transition proceeds by taking the system through an intermediate state made of non-linear triadic interactions within a set of discrete internal waves before a continuum of energy develops over the wave frequency domain and the wave vector space at the largest forcing amplitudes. 

In the most turbulent state, we observe $1$D vertical and horizontal spatial kinetic energy spectra compatible with power laws in $k_x^{-3}$ and $k_z^{-3}$. Such scalings have already been reported in several experimental and numerical studies of internal gravity wave turbulence~\cite{LeReun2018,Davis2020,Lanchon2023,Cortet2024}, where the $k^{-3}$ power laws were nevertheless not observed over more than half-a-decade of wave numbers. Here, we clearly observe this behavior over about a decade, at scales smaller than the injection scale, both in the horizontal and vertical directions. Such scaling in $k^{-3}$ are compatible with the raw prediction ---$1$D kinetic energy spectra in $N^2k^{-3}$--- that can be made on dimensional grounds for the regimes of ``saturated wave/critical balance'' turbulence in which the non-linear time saturates on the value of the wave period~\cite{Staquet2007,Nazarenko2011b,Cortet2024}. If this interpretation of our observations is correct, it would mean that the internal wave turbulence regime reached at high forcing amplitudes in our experiments is strongly non-linear and that the ratio of the Froude number to the non-dimensional wave frequency, $Fr/\omega_0^*\simeq 0.05$ (at the forcing scale), is still not sufficiently small to achieve a weakly non-linear wave turbulence.

Nevertheless, other observations presented in our study invite us to temper this interpretation in terms of strongly non-linear wave turbulence. We indeed show that the energy continuum observed at large forcing amplitudes is associated, in the spatio-temporal spectra, with a concentration of almost all the kinetic energy along the dispersion relation of internal gravity waves, as expected in a weakly non-linear regime. We also present in this article the bicoherence spectra of the measured velocity fields, an observable that probes the phase correlation within resonant triads of waves, which triads are precisely expected to drive the energy transfers in an internal wave turbulence~\cite{McComas1977,Dauxois2018}. For our experiment at the largest forcing amplitude, we report a bicoherence spectrum behaving as a relatively smooth function taking values of the order of the non-linearity parameter $Fr/\omega^*$ over the wave frequency domain. This observation and the fact that we observe a concentration of the energy on the dispersion relation in the spatio-temporal spectra, are in line with what is expected for an internal wave turbulence matching the classical theoretical assumptions of large domain and weak non-linearity ($Fr\ll\omega^*$)~\cite{Monsalve2020,Hasselmann1963}. In the event that this interpretation is correct and that the observed turbulence is not a strongly non-linear saturated-wave turbulence, another (weakly non-linear) interpretation should be made for the reported spatial kinetic energy spectra in $k^{-3}$. Unfortunately, we are unable to propose such an alternative theoretical description in the present article.

In any case, the question of whether the internal gravity wave turbulence we observe with clear power laws follows from a weakly or strongly non-linear dynamics remains open. A definitive experimental answer would involve exploring truly asymptotic regimes of stratified turbulence that would be both developed (high Reynolds number $Re$) and unambiguously weakly non-linear (low ratio of the Froude number to the non-dimensional wave frequency $Fr/\omega^*$). This implies increasing even further the injection wavelength and therefore also the size of the fluid domain to achieve, even better than in this work, the separation between the linear and non-linear time scales characteristic of weak non-linearity while keeping the flow turbulent.

\begin{acknowledgments}
This work was supported by a grant from the Simons Foundation (651461, PPC). We acknowledge J.~Amarni, A.~Aubertin, L.~Auffray, C.~Manquest, and R.~Pidoux for experimental help. 
\end{acknowledgments}

\appendix

\section{Quasi-steady mode}
\label{app:slowmode}

\begin{figure}[!htbp]
    \centering
    \includegraphics[width = 14cm]{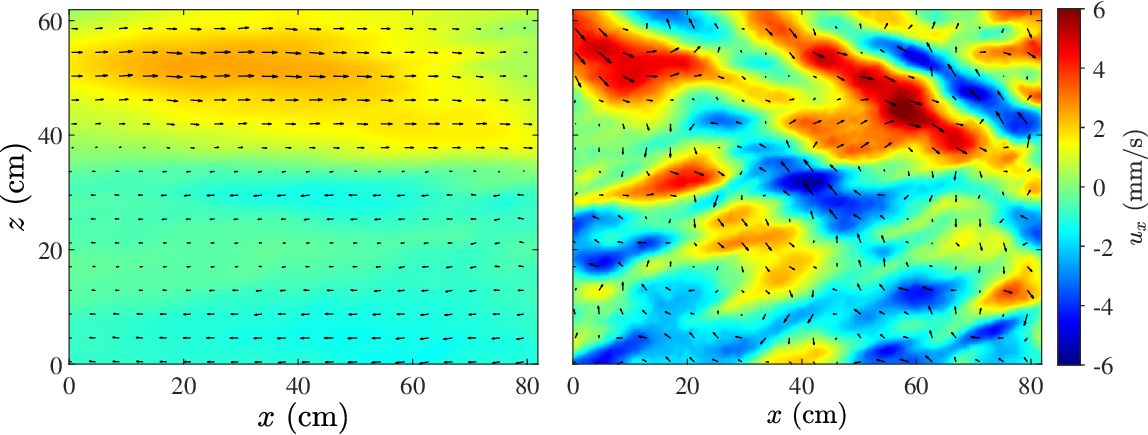}
    \caption{(Left panel) Velocity field of the quasi-steady mode of the experiment at $A=8$~mm computed by a temporal average of the measured velocity field over $100$ forcing periods during the statistically steady state of the experiment (between $t=900\,T$ and $1\,000\,T$ after the start of the wave generator; wide view camera). (Right panel) An instantaneous of the measured velocity field during the same stage of the flow for the same experiment (more precisely at $t = 964\,T$).}
    \label{fig:meanflow}
\end{figure}

In this appendix, we present, in the left panel of Fig.~\ref{fig:meanflow}, the velocity field of the quasi-steady mode for the experiment at $A=8$~mm. This field is computed by a temporal average of the measured velocity field over $100$ forcing periods during the statistically steady state of the experiment. For comparison, we show in the right panel of Fig.~\ref{fig:meanflow} a snapshot of the velocity field during the same stage of the flow.

\section{Spatio-temporal spectrum $E(\omega^*,k_x,k_z)$}
\label{app:spatiotemp}

\begin{figure}[!htbp]
    \centering
    \includegraphics[width=12cm]{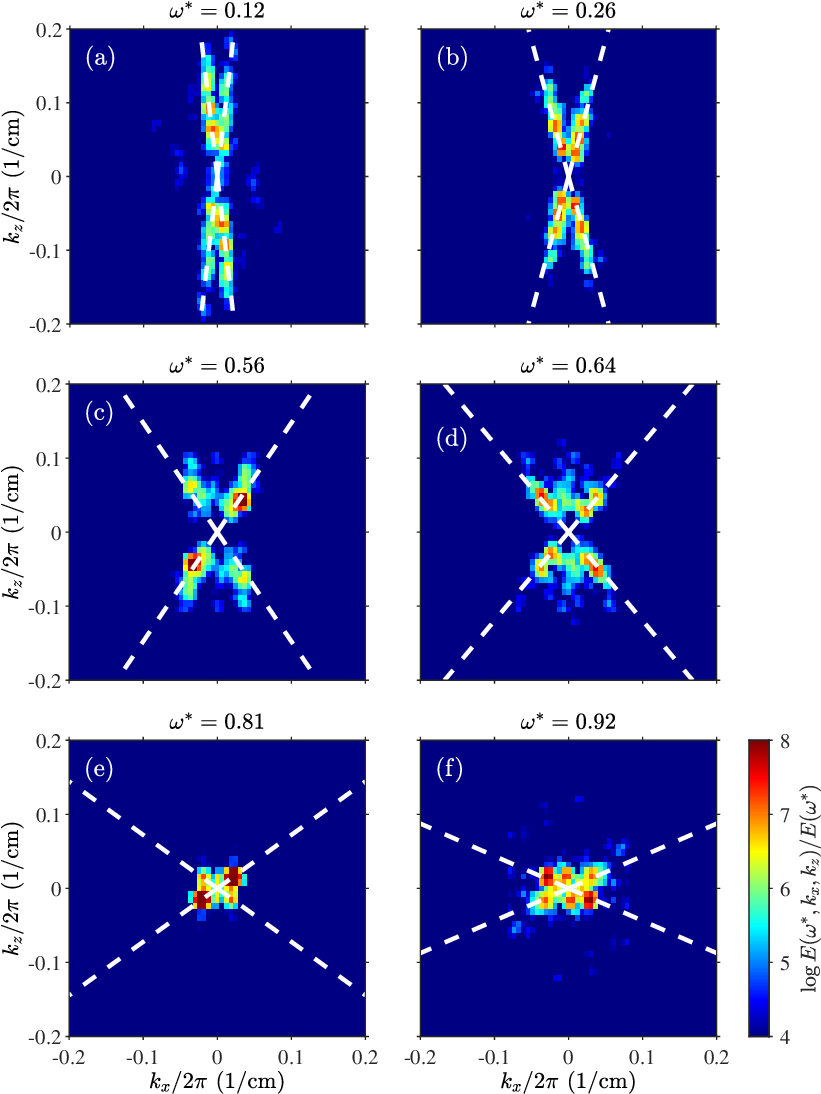}
    \caption{Logarithm of the normalized spatio-temporal kinetic energy spectrum $E(\omega^*,k_x,k_z)/E(\omega)$ for the experiment at $A=18$~mm for six values of the non-dimensional frequency, $\omega^*=0.12, 0.26, 0.56$, $0.64$, $0.81$, and $0.92$. In each panel, the dashed lines represent the dispersion relation $|k_z|=|k_x|(1/\omega^{*2}-1)^{1/2}$ of internal gravity waves at frequency $\omega^*$ and with $k_y=0$, i.e. propagating in the vertical measurement plane.}\label{fig:specSpatioTemp}
\end{figure}

In this appendix, we present examples of two-dimensional cuts of the spatio-temporal kinetic energy spectrum $E(\omega^*,k_x,k_z)$ from which the energy spectrum $E(\omega^*,\theta)$ as a function of the non-dimensional angular frequency $\omega^*$ and the angle  $\theta=\tan^{-1}(k_x/k_z)$, reported in Fig.~\ref{fig:omegaVsTheta}, is derived (see Eq.~\ref{eq:spectre_theta}). More precisely, we show in Fig.~\ref{fig:specSpatioTemp} the logarithm of the spatio-temporal spectrum normalized by the energy content at each frequency,  $\log[E(\omega^*,k_x,k_z)/E(\omega^*)]$, for the experiment at forcing amplitude ${A=18}$~mm for six frequencies in the internal wave domain, $\omega^*=0.12, 0.26, 0.56$, $0.64$, $0.81$, and $0.92$. The reader can find details on the calculation of $E(\omega^*,k_x,k_z)$ in Appendix~A of Ref.~\cite{Lanchon2023}.

In the panels of Fig.~\ref{fig:specSpatioTemp}, the dashed lines show the dispersion relation ${|k_z|=|k_x|(1/\omega^{*2}-1)^{1/2}}$ of internal gravity waves that are invariant in the $y$-direction, i.e., with $k_y=0$. In the general case, plane internal waves verify the dispersion relation $|k_z|=(k_x^2+ k_y^2)^{1/2}(1/\omega^{*2}-1)^{1/2}$ and will be associated in Fig.~\ref{fig:specSpatioTemp} to energy in the two regions defined by $|k_z|\geq |k_x|(1/\omega^{*2}-1)^{1/2}$. Nevertheless, it has been shown in Appendix~B of Ref.~\cite{Lanchon2023} that, even in the case of an ensemble of internal gravity waves with an axisymmetric distribution of wave vectors, we expect the spectrum $E(\omega^*,k_x,k_z)$ (computed from two-dimensional two-component PIV measurements in a vertical plane $y=y_0$) to be dominated by energetic spots close to the $2$D dispersion relation $|k_z|=|k_x|(1/\omega^{*2}-1)^{1/2}$, in a way similar to a flow composed only of waves propagating in the $(x,z)$ measurement plane ($k_y=0$).

In each panel of Fig.~\ref{fig:specSpatioTemp}, we see that most of the energetic regions are indeed found close to the $2$D dispersion relation ($k_y=0$), which observation is fully compatible with a flow composed of internal gravity waves verifying the dispersion relation. The energetic regions typically span a wave number range (in $k/2\pi$ units) from $0.02$ to $0.15$~cm$^{-1}$ which corresponds to lengthscales in the range from $7$ to $50$~cm. We also note a tendency for the energetic regions to spread toward lower scales for decreasing frequencies in agreement with the observation of Ref.~\cite{Lanchon2023}.


\end{document}